\documentclass[sigconf]{acmart}
\renewcommand\footnotetextcopyrightpermission[1]{} 
\usepackage{amsbsy,amsmath,amssymb,epsfig,hyperref,subcaption}
\usepackage[ruled, vlined]{algorithm2e}
\usepackage{subcaption}
\usepackage{multirow}
\usepackage{balance}
\usepackage{caption} 
\begin{document}
\title{Detecting Location Fraud in Indoor Mobile Crowdsensing}
\author{Qiang Xu}
\affiliation{%
	\institution{Department of Computing and Software\\ McMaster University}
	\streetaddress{1280 Main Street West}
	\city{Hamilton} 
	\state{Ontario}
	\country{Canada} 
	\postcode{L8S 4L8}
}
\email{xuq22@mcmaster.ca}

\author{Rong Zheng}
\affiliation{%
	\institution{Department of Computing and Software\\ McMaster University}
	\streetaddress{1280 Main Street West}
	\city{Hamilton} 
	\state{Ontario}
	\country{Canada} 
	\postcode{L8S 4L8}
}
\email{rzheng@mcmaster.ca}

\author{Ezzeldin Tahoun}
\affiliation{%
	\institution{Department of Computing and Software\\ McMaster University}
	\streetaddress{1280 Main Street West}
	\city{Hamilton} 
	\state{Ontario}
	\country{Canada} 
	\postcode{L8S 4L8}
}
\email{tahoune@mcmaster.ca}

\begin{abstract}
Mobile crowdsensing allows a large number of mobile devices to measure phenomena of common interests and form a body of knowledge about natural and social environments. In order to get location annotations for indoor mobile crowdsensing, reference tags are usually deployed which are susceptible to tampering and compromises by attackers. In this work, we consider three types of location-related attacks including tag forgery, tag misplacement and tag removal. Different detection algorithms are proposed to deal with these attacks. First, we introduce location-dependent fingerprints as supplementary information for better location identification. A truth discovery algorithm is then proposed to detect falsified data. Moreover, visiting patterns are utilized for the detection of tag misplacement and removal. Experiments on both crowdsensed and emulated dataset show that the proposed algorithms can detect all three types of attacks with high accuracy.  
\end{abstract}
\maketitle

\section{Introduction}
\label{sect:intro}
As a result of ubiquitous sensor peripherals and increasing computation capability of mobile devices, MCS (mobile crowdsensing), a special form of crowdsourcing where communities contribute sensing information and human intelligence using mobile devices, has shown great potential in a variety of environmental, commercial and social applications. Examples include measuring pollution levels in a city\cite{dutta2009common}, providing indoor localization~\cite{xuzheng16mobibee}, detecting potholes on a road~\cite{mohan2008nericell}, and many others. Among a variety of MCS applications, most of them require location tagging. Obtaining locations is trivial for outdoor applications since the Global Positioning System (GPS) complimented by cellular and WiFi access map based methods can provide accurate and robust location information in most outdoor environments. However, this is not the case for indoor applications. Despite the fact that people spend majority of their time indoor, indoor positioning systems (IPS) only have limited success due to the lack of pervasive infrastructural support, and the desire to keep user devices as simple as possible. Therefore, deploying a set of reference tags seems to be the best choice in order to provide accurate location tagging in indoor MCS applications.

As an example, reference tags are instrumental in collecting crowdsensed data for fingerprint-based indoor localization. Generally, fingerprint-based localization works in two stages: training and operational stages. In the training stage, comprehensive site survey is periodically conducted to record the fingerprints at targeted locations. In the operational stage, when a user submits a location query with his current fingerprints, a localization server computes and returns the estimated location. Comprehensive site survey is time-consuming, labor-intensive, and subjects to environmental changes. This makes it a perfect candidate application for mobile crowdsensing. Specifically, we can deploy a certain number of reference tags at targeted locations and then recruit volunteers or workers to contribute fingerprints at these tags.

\begin{figure}[tbp]
	\centering
	\includegraphics[width=0.4\textwidth]{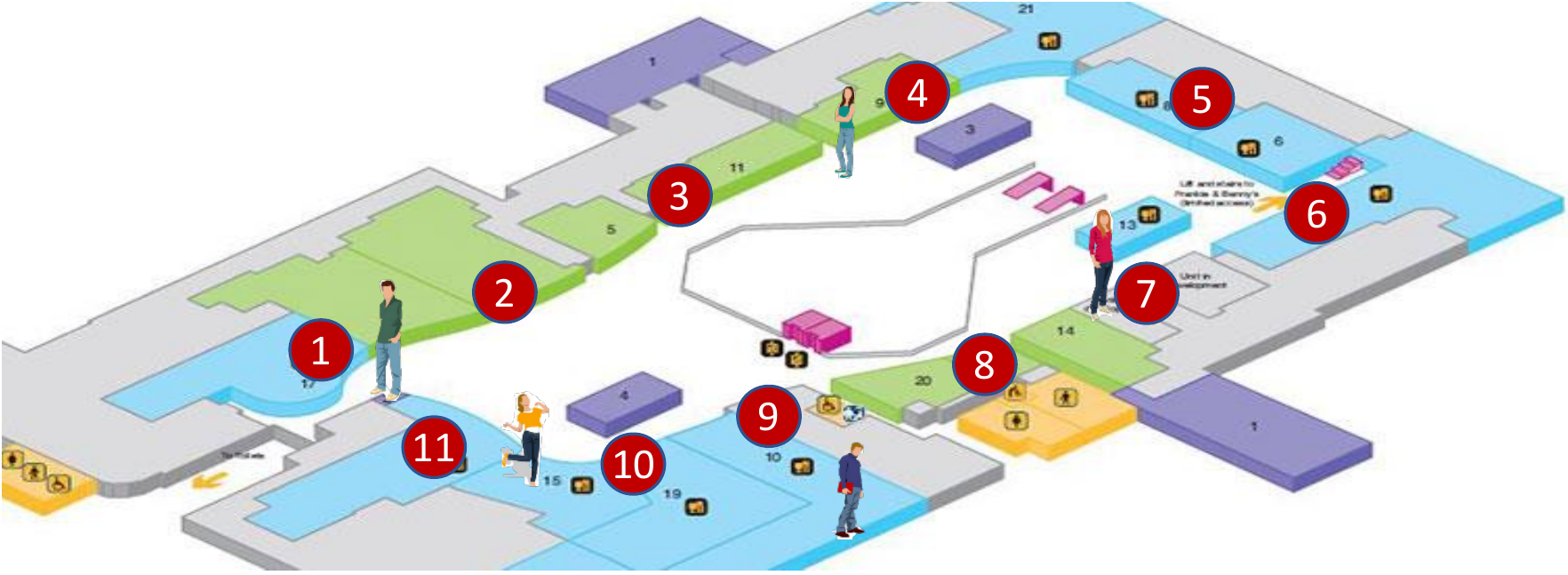}
	\caption{An example of using reference tags for indoor mobile crowdsensing. The users are required to upload sensing data at these tags (red numbered dots).}
	\label{fig:demo}
\end{figure}

However, the adoption of reference tag based method is not without cost. It faces the challenges of incurring additional costs in deploying the tags and malicious behaviors from users. Tags such as QR codes, though cost next to nothing, are susceptible to duplication, damage and removal. As a result, the crowd-sensed data can be of low quality. For example, a user can easily copy a QR code and scan it at somewhere else when QR codes are used as tags. In our previous experience on MobiBee, a game for mobile crowdsensing, we observed cheats during the data collection\cite{xuzheng16mobibee}. In machine learning, it is well known that ``garbage in garbage out''. Using crowdsensing data with contaminated location tags will necessarily lead to poor trained models. Thus, it is important to safe guard against fraudulent behaviors and falsified data, especially when incentives are involved. 

In this work, we consider three types of attacks for location-tagged mobile crowdsensed data:
\begin{enumerate}
\item \textbf{Tag Forgery}: An attacker forges one or multiple tags and uses them to annotate the collected data.
\item \textbf{Tag Misplacement}: A tag is moved from its designated place to somewhere else.
\item \textbf{Tag Removal}: A tag is removed illegally.
\end{enumerate}

We believe that most of the location-related malicious behaviors can be represented by these three types of attacks. Different detection algorithms are proposed to deal with these attacks. First, as supplementary information, location-dependent fingerprints are introduced for better location identification. A truth discovery algorithm is then proposed to detect falsified data. Moreover, visiting patterns are utilized for the detection of tag misplacement and removal. Experiments on both crowdsensed and emulated dataset show that the proposed algorithms can detect all three types of attacks with high accuracy.

The rest of the paper is organized as follows. First, we provide a review of the related work in Section~\ref{sect:related}. The three types of attacks are elaborated in Section~\ref{sect:threeattack}. Attack detection algorithms are presented in Section~\ref{sect:cheat_detection}.  Experimental results are given in Section~\ref{sect:exp}. Finally, we conclude the paper with summary and future work.

\section{Related work}
\label{sect:related}
There has been some work on leveraging mobile crowdsensing for indoor applications. With the booming research on indoor localization, many indoor positioning systems try to take advantage of MCS. A comprehensive review can be found in \cite{pei2016survey}. Generally, these applications fall into two broad categories based on the purpose: for fingerprinting and for map construction. In the first category, schemes such as Zee and LiFS rely on an alternative odometry based solutions for initial localization or displacement estimates\cite{Rai2012,Yang2012}. MobiBee is a crowdsensing solution for location fingerprint collection that participants are uncontrolled and utilize their own devices~\cite{xuzheng16mobibee}. Several schemes have been proposed in literatures to construct indoor maps by organically combining users' trajectories through pedestrian dead reckoning~\cite{gao2014jigsaw, Shen2013WIP}. These schemes differ in how the trajectories are combined. In \cite{chen2015rise}, the authors propose IndoorCrowd2D to reconstruct the indoor interior view of a building from crowdsensed data.

Despite many work on indoor positioning system, deploying reference tags seems to be the simplest to implement. A variety of techniques, including Radio Frequency Identification (RFID), Near Field Communication (NFC), Quick Response (QR) code, have been exploited for location reference~\cite{bouet2008rfid,ni2004landmarc,siira2009location,xu2016mobibee,lee2012qrloc}. Though each of these techniques has its own advantages, none of them are absolute safe in terms of is nonforgeability. Security threats in NFC and RFID have been studied in~\cite{kim2013study, langheinrich2009survey}. QR codes can be easily copied and spread because they not have any copy protection mechanism. QR code forgery attack has been reported in our previous work\cite{xu2016mobibee}. In order to provide verifications for users' location claims, location proof mechanisms have been proposed~\cite{khan2014and, sastry2003secure}. However, these methods require either additional infrastructures or lots of users which will complicate the design of a mobile crowdsensing system. 

Truth discovery is a process to integrate multi-source noisy information by estimating the reliability of each source. A comprehensive survey on truth discovery can be found at~\cite{li2016survey}. Generally, a truth discovery task can be modeled in three ways: iterative methods\cite{dong2009integrating}, optimization based methods\cite{aydin2014crowdsourcing}, and probabilistic graphical model based methods\cite{zhao2012bayesian}. In this work, we formulate the falsified data detection as a truth discovery problem and propose a probabilistic graphical model.

\section{Attacks on Location Tags}
\label{sect:threeattack}
There are many security threats that can compromise the data quality while conducting mobile crowdsensing. In this work, we only target those that can affect the location tags. The crowdsensed data with location tags can be described as $\langle rid, uid, lid, sensed\_data \rangle$, where $rid$ is the record index, $uid$ is the user ID, $lid$ is the location tag ID, and $sensed\_data$ is the sensed data. Take the application of surveying the temperature of a building as an example, we can deploy a set of reference tags and ask workers to contribute the temperatures at these tags. Table~\ref{tab:probdemo} is a piece of the crowdsensed data. We categorize these location-related attacks into three types, i.e., \textit{tag forgery}, \textit{tag misplacement}, and \textit{tag removal}. The rationale behind these attacks will be discussed in the rest of this section. 

%

\subsection{Tag Forgery Attack}
The first and primary type of attack is called \textit{tag forgery}. During the operation of MCS, an attacker can forge one or multiple location tags and use these illegitimate tags to annotate the data. As a result, the location tags will be compromised. This type of attack will introduce falsified data and the contributors of these falsified data can be considered as attackers. \textit{Tag forgery} can be motivated by profits or intentions to sabotage. Often times it is easier to generate fraudulent data than legitimate ones as the later requires a participant to physically travel to target locations.

\subsection{Tag Misplacement Attack}
Another type of attack is \textit{tag misplacement}. After the deployment of reference tags, some tags may be moved from their designated places to somewhere else. For example, a tag could be moved by the custodian if it blocks important signages, or a tag might be intentionally moved to another place by malicious users. Similar to \textit{tag forgery} attack, \textit{tag misplacement} will introduce falsified data as well. However, worse than \textit{tag forgery}, all subsequent data will be falsified if a tag has been moved. Therefore an additional detection mechanism is required to enable the administrator to correct the misplaced tags in time.

\subsection{Tag Removal Attack} 
In \textit{tag removal}, a tag may be removed by accidence or malicious behaviors. In this case, there will be no falsified data. However, this type of attacks will lead to an incomplete dataset. In the aforementioned temperature survey example, we might never get the temperatures at some target locations due to \textit{tag removal} attack. Therefore, a detection method would be helpful to inform the administrator to redeploy the missing tags. 

In summary, both \textit{tag forgery} and \textit{tag misplacement} can introduce falsified data, and hence a falsified data detection algorithm is needed. As for \textit{tag misplacement} and \textit{tag removal}, actions are required when the attacks happen. The details of the corresponding detection algorithms will be elaborated in the following section.

\section{Attack Detection}
\label{sect:cheat_detection}
If the data to be collected contains little location dependent information, it is not possible to detect different forms of attacks. To make our detection algorithms agnostic to individual applications, we take advantage of the location-dependent fingerprints. We collect these fingerprints in addition to the required data during the operation of MCS. These fingerprints serve as additional identification of location. Location-dependent fingerprints have been well studied for indoor localization. Classical fingerprints include WiFi received signal strength (RSS), magnetic field magnitude, illumination level etc. In this work, we use WiFi RSS and manetic field magnitude. Note that unlike fingerprinting based localization, we do not perform site survey to construct fingerprint maps. The problem can therefore be described as: \textbf{given a dataset in the form of $\langle rid, uid, lid, sensed\_data, fingerprint\rangle$, how to detect the three types of attacks?} Table~\ref{tab:probdemo} is a demonstration of this attack detection problem. It is not trivial due to the absence of ground-truth fingerprint and sensing data at each location.
\begin{table}[tbp]
	\centering
	\small
	\caption{A demonstration of the formulated attack detection problem}
	\label{tab:probdemo}
	\begin{tabular}{| c | c | c | c | c | c | p{1cm} |}
		\hline
		\multicolumn{6}{|c|}{Raw Crowdsensed Data} & Output\\
		\hline
		\textbf{rid} & \textbf{lid} & \textbf{uid} & \textbf{sensed\_data} & \textbf{FP}& \textbf{Time-stamp} & \textbf{location validity}\\
		\hline
		1 & $loc_1$ & $u_1$ & $30^\circ $C & $fp_1$&2017-1-09 11:20:20 &1\\
		\hline
		2 & $loc_2$ & $u_2$ & $31^\circ $C & $fp_2$&2017-1-09 12:10:20 & 0\\ 
		\hline
		$\vdots$ & $\vdots$ & $\vdots$ & $\vdots$& $\vdots$& $\vdots$ &$\vdots$\\ 
		\hline
	\end{tabular}
\end{table}

\subsection{Detection of falsified data}
\label{sect:falsifieddetection}
\subsubsection{Coarse-grained detection}
Under the \textit{tag forgery} and \textit{tag misplacement} attacks, the required data along with the fingerprints are collected at illegitimate locations. Comparing with the reference tag, the fingerprints are much more difficult to be forged. Therefore, it is possible to filter the contributed data by simply checking the validity of the associated fingerprints. Specifically, we can check the service set identifications (SSIDs) in the fingerprints to detect problematic scans that do not contain the legitimate SSIDs. 

In addition to checking the validity of SSIDs, we can further detect the anomalies based on the time stamps of the contributed data and the topology information of all location tags. The idea is that a user needs to travel from one location to another to collect data. Therefore, we can estimate the user's walking speed based on the time stamp and topology information of location tags. Under the \textit{tag forgery} attack, the attackers do not have to ``travel'' physically. For example, an attacker may forge all the location tags and collect data from the same location but tags them with different locations. A speed anomaly might be detected if this is the case. Formally, we use  $\langle \mathbf{X}^{u}_i,  \mathbf{X}^u_{i+1} \rangle$ to denote two consecutive uploads contributed by user $u$, where $\mathbf{X}^{u}_i = (loc^{u}_i, ts^u_i, \cdots)$, $\mathbf{X}^{u}_{i+1} = (loc^{u}_{i+1}, ts^u_{i+1}, \cdots)$, where $loc_i^u$ represents the location tag and $t_i^u$ is the time stamp. We flag both  $ \mathbf{X}^{u}_i$ and  $\mathbf{X}^u_{i+1}$ as falsified if $\frac{dist(loc^{u}_{i+1}, loc^{u}_{i})}{ts^u_{i+1} - ts^u_i} > \rho$ where $\rho$ is the maximum of human walking/running speed. In our experiments, we set $\rho = 10m/s$.

Though both SSIDs checking and speed anomaly detection can benefit \textit{tag forgery} detection, given the small collection of SSIDs and the ease of changing SSIDs on WiFi APs, such solutions are likely to work only with unsophisticated perpetrators. There is a need to develop more generic detection algorithms.

\subsubsection{Truth-discovery based detection}
\label{sect:truthdiscovery}
Though it is reasonable to assume majority of the contributors are honest, one cannot assume the majority of the collected data is legitimate. Furthermore, it may be insufficient to identify the ``bad guy" as the bad guy may still contribute ``good" data. For instance, a perpetrator scans legitimate tag once before commencing on tag forgery attacks since she has to explore the legitimate tags anyway. Under these considerations, we formulate the attack detection problem as a truth discovery problem. Truth discovery integrates multi-source noisy information by estimating the reliability of each source. Interested readers can refer to \cite{li2016survey} for a comprehensive survey on truth discovery. 

\begin{figure}[tbp]
	\centering
	\includegraphics[width=0.25\textwidth]{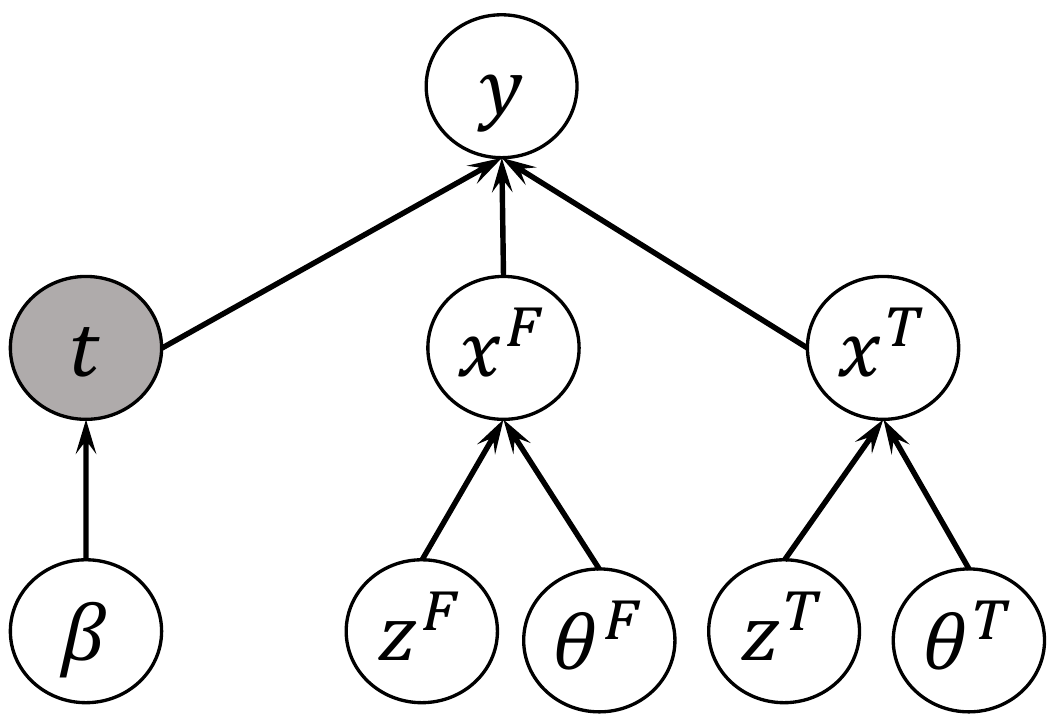}
	\caption{The probabilistic graphical model}
	\label{fig:graphic}
\end{figure}

We first describe the notations used and then give a formal definition of the truth discovery problem.

\paragraph{\textbf{Input}} Given a set of locations: $\mathcal{K} = \{k\}_{k=1}^K$ that we are interested in, and a set of users (the community) $\mathcal{U} = \{u\}_{u=1}^U$ which provide sensing data along with fingerprints from all $\mathcal{K}$ locations. The raw dataset is denoted as $\mathcal{X} = \{(\mathbf{x}_1, u_1, z_1, ts_1),\cdots, (\mathbf{x}_N, u_N, z_N, ts_N)\}$, where $\mathbf{x}_i$ is the fingerprints (the sensing data can be included if it is location dependent), $z_i \in \mathcal{K}$ is the location index, $u_i \in \mathcal{U}$ is the user index, and $ts_i$ is the time stamp.

\paragraph{\textbf{Output}} The location validity labels for the given dataset are denoted as $\mathbf{t} = (t_1,\cdots, t_N)$, where
\begin{equation}
t_i = 
\begin{cases}
& 1 \   \text{if the location tag $\mathbf{z}_i$ is truthful}\\
& 0 \   \text{if the location tag of $\mathbf{z}_i$ is falsified.}
\end{cases}
\end{equation}

\paragraph{\textbf{Truth Discovery Task}}
The truth discovery task is formally defined as estimating $\mu$ and $\mathbf{t}$ given the raw dataset $\mathcal{X}$.

In order to tackle this problem, we propose a probabilistic graphical model, as illustrated in Figure~\ref{fig:graphic}. Specifically, we use two Gaussian Mixture Models (GMM) $p(\mathbf{x}|\theta^T)$ and $p(\mathbf{x}|\theta^F)$ to model the truthful and falsified fingerprints, respectively. For fingerprints with truthful location tags, each component corresponds to a unique location. As for fingerprints with falsified location tags, we assume that all these fingerprints from the same user are collected at the same illegal location. Hence, each component in the GMM of falsified data corresponds to a unique user. As such, the total number of components for $p(\mathbf{x}|\theta^T)$ is $K$, and the total number of components for $p(\mathbf{x}|\theta^F)$ is $U$. The two models are denoted as
\begin{align}
p(\mathbf{x}|\theta^T) &= \sum_{k=1}^{K}\alpha^T_k p(\mathbf{x}| \theta^T_k)\\
p(\mathbf{x}|\theta^F) &= \sum_{u=1}^{U}\alpha^F_u p(\mathbf{x}| \theta^F_u)
\label{eq:gmm}
\end{align}

with mixing weights that $\sum_k\alpha^T_k = 1$ and $\sum_u\alpha^F_u = 1$, $\theta^T_k = (\mu^T_k, \Sigma^T_k)$, and $\theta^F_u = (\mu^F_u, \Sigma^F_u)$. Unlike the traditional EM algorithm for GMM where the component indicator is an unknown hidden variable, both $z_i$ and $u_i$ are known in our case. In our model, the location validity variable $t_i$ is considered as hidden variable that needs to be estimated. 

The associated log-likelihood for the data $\mathbf{x}_i$ with $i \in \{ 1, \cdots, N \}$ is given as $ \mathcal{L}(\Theta) = \sum_{i}\log \sum_{t_i}p(\mathbf{x}_i, t_i;\Theta)$ where $\Theta = (\theta^T, \theta^F, \beta)$, $\beta$ represents user reliability which will be elaborated later.

For each $i$, let $Q_i$ be some distribution over the $t_i$'s($\sum_{t_i} Q_i(t_i) = 1$, $Q_i(t_i) \geq 0$), then we have
{\small
\begin{align}
\mathcal{L}(\theta) & = \sum_i\log \sum_{t_i} Q_i(t_i) \frac{p(\mathbf{x}_i, t_i; \Theta)}{Q_i(t_i)} \geq \sum_i \sum_{t_i} Q_i(t_i)\log \frac{p(\mathbf{x}_i, t_i; \Theta)}{Q_i(t_i)}
\end{align}
}

In such a setting, the EM algorithm gives us an efficient method for maximum likelihood estimation. Instead of maximizing $\mathcal{L}(\Theta)$ directly, we repeatedly construct a lower bound of $\mathcal{L}(\Theta)$ (E-step) and optimize that lower bound (M-step). We can now write down the sequence of our EM algorithm for our truth discovery task:
\begin{align}
E-step: & \text{ For each $i$, set }\\
Q_i(t_i) & = p(t_i|\mathbf{x}_i, \theta^T, \theta^F, \beta) = \frac{p(\mathbf{x}_i|t_i, \theta^T, \theta^F) \cdot p(t_i|\beta)}{C}\\
M-step:&\text{ Set}\\ 
\Theta & = \arg\max_\Theta \sum_i \sum_{t_i} Q_i(t_i) \log \frac{p(\mathbf{x}_i, t_i; \theta)}{Q_i(t_i)}
\end{align}
where $\beta_u$ is the reliability of user $u$ and $\beta_u = 1$ if all the the data from user $u$ are completely truthful, $p(t_i=1|\beta) = \beta_{u_i}$, and $C$ is a constant. Let us define
\begin{equation}
J(Q, \Theta) = \sum_i \sum_{t_i} Q_i(t_i) \log \frac{p(\mathbf{x}_i, t_i; \Theta)}{Q_i(t_i)}
\end{equation}
Then we can maximize $J(Q, \Theta)$ with respect to $\theta^T$, $\theta^F$, and $\beta$ respectively,
{\small
\begin{align}
\frac{\partial J(Q, \Theta)}{\partial \theta_k^T} & = \sum_{i:z_i = k} Q_i(t_i = 1) \frac{\frac{\partial}{\partial \theta_k^T} p(\mathbf{x}_i|\theta_k^T)}{p(\mathbf{x}_i|\theta_k^T)} = 0\\
\frac{\partial J(Q, \Theta)}{\partial \theta_u^F} & = \sum_{i:u_i = u} Q_i(t_i = 0) \frac{\frac{\partial}{\partial \theta_u^F} p(\mathbf{x}_i|\theta_u^F)}{p(\mathbf{x}_i|\theta_u^F)}= 0\\
\frac{\partial J(Q, \Theta)}{\partial \beta_u} &= 0
\end{align}
}
After straightforward derivations, we can update $\theta^T$, $\theta^F$ and $\beta$ in M-step as:
{\small
\begin{align}
\alpha_k^T &= \frac{\sum_{i:z_i=k}Q_i(t_i=1)}{\sum_i Q_i(t_i=1)}\\
\mu_k^T & = \frac{\sum_{i:z_i=k} Q_i(t_i=1)\mathbf{x}_i}{\sum_{i:z_i=k} Q_i(t_i=1)}\\
\Sigma_k^T & = \frac{\sum_{i:z_i=k} Q_i(t_i=1)(\mathbf{x}_i - \mu_k^T)(\mathbf{x}_i - \mu_k^T)^\top }{\sum_{i:z_i=k} Q_i(t_i=1)}\\
\alpha_u^F &= \frac{\sum_{i:u_i=u}Q_i(t_i=0)}{\sum_i Q_i(t_i=0) }\\
\mu_u^F & = \frac{\sum_{i:u_i=u} Q_i(t_i=0)\mathbf{x}_i}{\sum_{i:u_i=u} Q_i(t_i=0)}\\
\Sigma_u^F & = \frac{\sum_{i:u_i=u} Q_i(t_i=0)(\mathbf{x}_i - \mu_u^F)(\mathbf{x}_i - \mu_u^F)^\top }{\sum_{i:u_i=u} Q_i(t_i=0)}\\
\beta_u & = \frac{\sum_{i:u_i = u} Q_i(t_i=1)}{\sum_{i:u_i = u}1}
\end{align}
}

With these, we can estimate the hidden variable $t_i$ repeatedly until convergence. Similar to standard EM, the convergence of our algorithm can be proved by showing that our choice of the $Q_i$'s always monotonically improves the log-likelihood. In the final falsified data detection algorithm, the coarse-grained method is performed in the beginning and the result is used as the initial value of $\mathbf{t}$.

\subsection{Detecting tag misplacement}
The detection of \textit{tag misplacement} is based on the visiting pattern of location tags. In indoor MCS, it is likely that a user will contribute sensing data in a continuous manner. All sensing data uploaded from user $u$ can be sorted in chronological order and represented as $traj^u = \langle(loc^u_1,ts^u_1), (loc^u_2, ts^u_2), \cdots \rangle$. We define a length-$m$ trajectory of user $u$ as $traj_{m,i}^u=\langle (loc^u_i,ts^u_i), \cdots, (loc^u_{i+m-1},ts^u_{i+m-1}) \rangle$ where $traj_{m,i}^u$ is a subsequence of $traj^u$ and $ts_{j+1} - ts_j \leq TW $ for any $i \leq j \leq i+m-2$, $TW$ is a time difference threshold. 

\begin{figure}[tbp]
	\centering
	\includegraphics[width=0.4\textwidth]{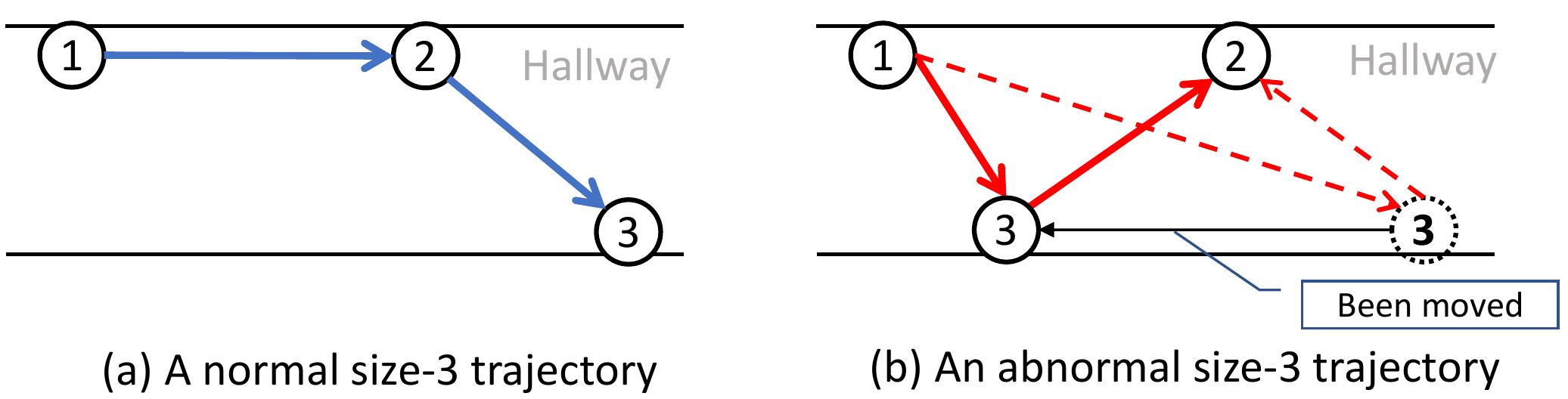}
	\caption{An example of normal and abnormal length-3 trajectories. In (b),  tag 3 has been moved from dotted circle to solid circle.}
	\label{fig:size3traj}
\end{figure}

Given $TW$, the raw data set $\mathcal{X} = \{\mathbf{X}_1, \cdots, \mathbf{X}_N\}$ can be easily transformed into a set of length-3 trajectories which can be used for \textit{tag misplacement} detection. We introduce the \textit{normal} and \textit{abnormal} length-3 trajectories. A length-3 trajectory $traj_{i}^{u,3}=\langle (loc^u_i,ts^u_i), (loc^u_{i+1},ts^u_{i+1}), (loc^u_{i+2},ts^u_{i+2})\rangle$ is normal if $dist(loc^u_{i+2}, loc^u_i) \geq dist(loc^u_{i+1}, loc^u_i)$, and abnormal otherwise. Figure~\ref{fig:size3traj} provides an example of normal and abnormal length-3 trajectories. Although an occasional abnormal length-3 trajectory does not necessarily mean \textit{tag misplacement}, frequent abnormal length-3 trajectories around a location tag usually implies that this tag has been moved. Based on this observation, a threshold based \textit{tag misplacement} detection algorithm is described as Algorithm~\ref{alg:misp}.

\begin{algorithm}
	\small
	\KwIn{Raw dataset: $\mathcal{X}$}
	\KwOut{A set of misplaced location tags}
	Initialize each $count_{loc_k}$ to 0\;
	$traj^3$ = a set of length-3 trajectories generated from $\mathcal{X}$\;
	\For{ $traj^{3,u}_i \equiv  \langle (loc^u_i,ts^u_i), (loc^u_{i+1},ts^u_{i+1}), (loc^u_{i+2},ts^u_{i+2})\rangle\in traj^3$}
	{ 
	\If{$traj^3_i$ is abnormal}{
		$count_{loc_{i+1}} = count_{loc_{i+1}} + 1$\;
	}

	Sort $\langle loc_1, \cdots, loc_K \rangle$ in descending order based on $count$\;
	Flag the top-ranking $loc_k$s as misplaced\;	
}

\caption{\textit{tag misplacement} detection}
\label{alg:misp}
\end{algorithm}

\subsection{Detection of tag removal}
\label{sect:detectremoval}
In real world, a reference tag could be removed on purpose or by accidence. Once a tag has been removed, it will never show up in the rest of data collection except that it might be forged by some attackers. The detection of \textit{tag removal} relies on the observation that a tag could have been removed if it does show up but its adjacent tags show up frequently. Based on this observation, the \textit{tag removal} attack can be detected by comparing the frequency of a tag with the frequencies of its neighbors. Concretely, all location tags can be ranked in ascending order based on $\frac{freq_{loc_k}}{mean(\{freq_{neighbor^k_1}, freq_{neighbor^k_2}, \cdots\})}$, where $dist(neighbor^k_j, loc_k) \leq \phi$. The distance threshold $\phi$ is set to 5m in our experiment. Then the top-ranking location tags can be flagged as removed. In practice, this detection method can be easily converted to an online algorithm by setting a threshold.

\section{Evaluation and Results}
\label{sect:exp}
\subsection{Two datasets}
The experiments are based on both crowdsensed and emulated dataset. The real world dataset is collected by a treasure hunt game MobiBee which is designed for location dependent fingerprint collection. QR codes are used as location tags. Various incentive strategies, including monetary, entertainment and competition, are utilized for better user participation. More details can be found in~\cite{xuzheng16mobibee} about the game design. Over the one month operation, 26 users have signed up the game but only 14 of them actually contribute data. In the experiments, we select 5 users who contribute the vast majority of the data. During the operation of MobiBee, two users have been found cheating which can be categorized as \textit{tag forgery} attack. However, both \textit{tag misplacement} and \textit{tag removal} did not appear in the data collection of MobiBee. Therefore, another dataset is needed to evaluate the performances of proposed \textit{tag misplacement} and \textit{tag removal} detection algorithms. 

The second dataset is constructed by emulation. Concretely, we develop a fingerprint collection app. Several volunteers are recruited to travel around inside the target building and upload sensing data at a set of pre-selected locations. The trajectories of these volunteers are carefully designed to imitate the actual scenario of MCS. 20 users have been emulated and each user has been to every location at least once. 5 users have been designed as attackers who perform \textit{tag forgery} attack. 5 location tags are moved to illegitimate places during the data collection, as the result of \textit{tag misplacement} attack. Similarly, another 5 location tags are removed at some points, as the result of \textit{tag removal}. The details of the two datasets are given in Table~\ref{tab:twodatasets}.

\begin{table}
	\centering
	\small
	\caption{The details of two datasets}
	\label{tab:twodatasets}
	\begin{tabular}{| c | c | c |}
		\hline
		& MobiBee Dataset & Emulated Dataset\\
		\hline
		\# of location tag & 24 & 50\\
		\hline
		\# of user & 5 & 20\\
		\hline
		\# of attacker & 2 & 5\\
		\hline
		\# of record & 21722 & 17487\\
		\hline
		\# of falsified record & 8338& 3186\\
		\hline
		\# of misplaced tags & 0 & 5\\
		\hline
		\# of removed tags & 0 & 5\\
		\hline
	\end{tabular}

\end{table}

\subsection{Performance of falsified data detection}
\begin{figure}[tbp]
	\centering
	\includegraphics[width=0.5\textwidth]{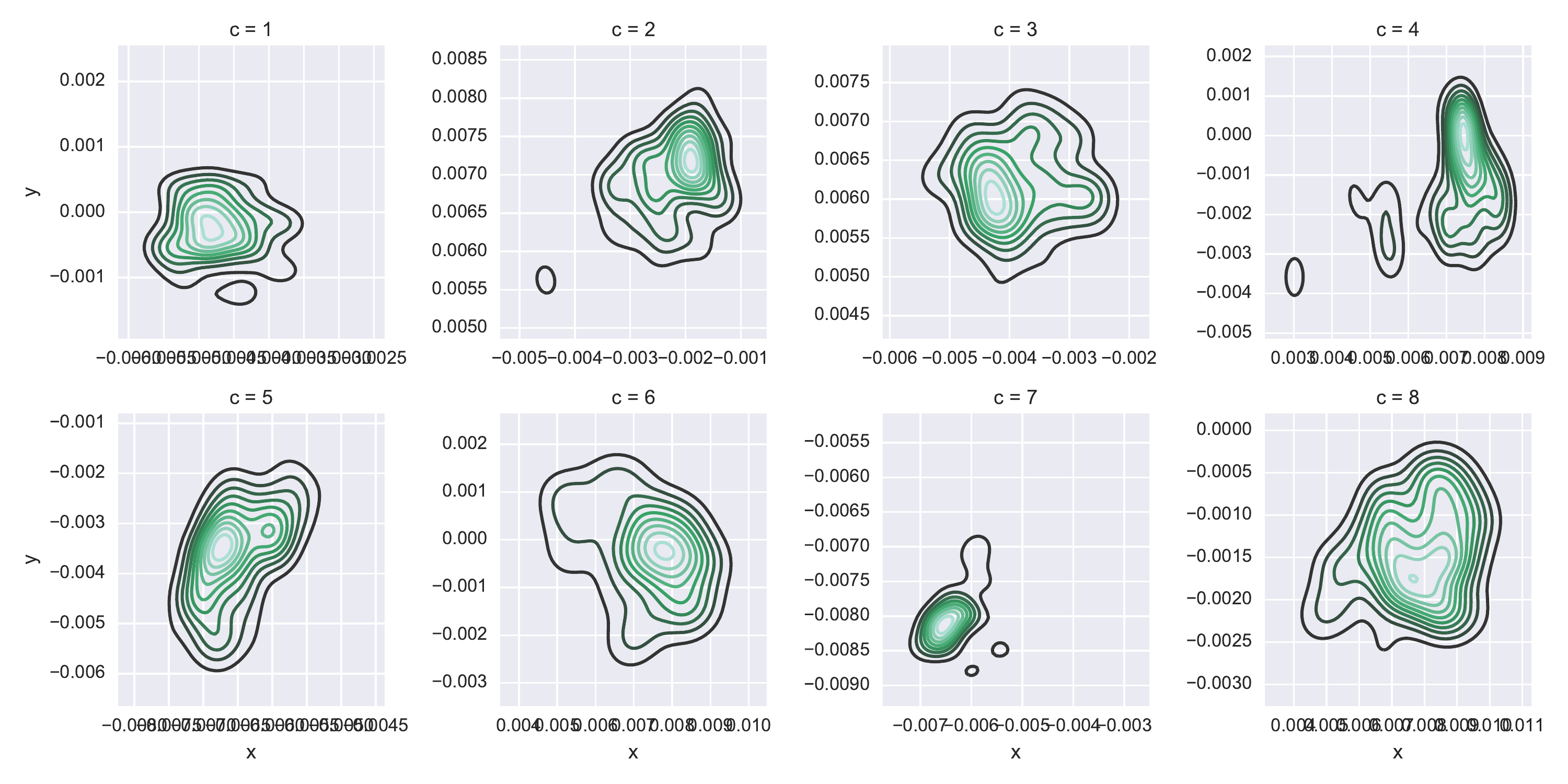}
	\caption{Gaussian kernel density estimation of fingerprints collected at different locations}
	\label{fig:kdemobibee}
\end{figure}

In section~\ref{sect:falsifieddetection}, we model the fingerprints collected from different locations as Gaussian Mixture. The Gaussianality of fingerprints collected from the same location is demonstrated as Figure~\ref{fig:kdemobibee}.

Two users have been found cheating in MobiBee. The falsified instances in MobiBee dataset were first detected by SSID checks and are further confirmed by interviewing the respective perpetrators, and thus the verified data is used as ground truth. Therefore, the coarse-grained detection methods are not used for MobiBee dataset. As for the emulated dataset, it is a different story since we know the ground truth in the first place. Therefore, both coarse-grained detection algorithm and truth discovery based algorithm are applied on the emulated dataset. The experiment results show that the proposed algorithm can successfully detect falsified data with high accuracy. The detailed results of the experiments on two different datasets are reported as Table~\ref{tab:results}. Note that, the recall of the coarse-grained detection algorithm is very low whereas the precision is very high, which means that although coarse-grained detection approach is accurate it is not capable enough to capture all falsified data. After integrating the truth discovery algorithm, we can see that the recall has been significantly improved despite that the precision is sacrificed. While detecting falsified data, False Negative (FN) error is more harmful than False Positive error (FP), since it can contaminate the dataset. In this sense, combining coarse-grained and truth discovery algorithm is more desirable.

\begin{table}
	\centering
	\small
	\caption{The performance of falsified data detection algorithm}
	\label{tab:results}
	\begin{tabular}{| p{2.5cm} | p{1cm} | p{2cm} | p{2cm} |}
		\hline
		&MobiBee Dataset & Emulated Dataset (coarse-grained) & Emulated Dataset (Integrated)\\
		\hline
		Overall Accuracy & 88.4\%& 83.4\% & 87.2\%\\
		\hline
		Precision for Falsified Data & 87.2\%& 100\%&62\%\\
		\hline
		\textbf{Recall for Falsified Data}& \textbf{81.7\%} & \textbf{8.9\%}&
		\textbf{77.6\%}\\
		\hline
	\end{tabular}
\end{table}

\subsection{Estimation of user reliability $\beta$}
A side output of the truth discovery algorithm in Section~\ref{sect:truthdiscovery} is user reliability $\beta$. For MobiBee dataset, there are 5 users in total and 2 of them were found cheating. After the convergence of the algorithm, the estimated $\beta_{1-5}$ are listed in the first row of Table~\ref{tab:user}. This estimation is consistent to the ground truth that user 2 and 5 are attackers. While using the emulated dataset, the estimated user reliabilities are provided in the rest of Table~\ref{tab:user}. We can easily observe that user 1-10 have relatively worse reliability than others. This accurately reflects the fact that user 1-5 upload some data from misplaced tags, and user 6-10 are attackers who upload some data using forged tags. 
\begin{table}[tbp]
	\centering
	\small
	\caption{The estimated user reliability}
	\label{tab:user}
	\begin{tabular}{| c | c |}
		\hline
		 & Estimation\\
		\hline
		MobiBee DS&\\
		$\beta_{1-5}$ & [0.97129253,  0.32158027, 1., 0.96290124, 0.522175211]\\
		\hline
		Simulated DS&\\
		$\beta_{1-5}$& [0.5121315 ,  0.67818931,  0.66130878,  0.61648436,  0.62845224]\\
		$\beta_{6-10}$& [0.33275762,  0.46122064,  0.42303993,  0.46608656,  0.30396718]\\
		$\beta_{11-15}$&[0.91738818,  0.95683221,  0.93786919,  0.97727463,  1.0]\\
		$\beta_{16-20}$&[0.97310623,  0.93121491,  0.98167733,  0.99308929,  0.96049801]\\
		\hline
	\end{tabular}
\end{table}

\subsection{Performance of tag misplacement detection}
According to Table~\ref{tab:twodatasets}, only the emulated dataset contains \textit{tag misplacement} attack and hence is used to evaluate the \textit{tag misplacement} detection performance. Tag $\{5, 15, 25, 35, 45\}$ were actually moved during the data collection. Based on Algorithm~\ref{alg:misp}, the top-10 candidates of misplaced tags are $\{5, 15, 25, 45, 29, 12, 33, 35, 20, 14\}$, which capture all 5 misplaced location tags. We will capture 4 out of 5 misplaced tags if we flag top-5 candidates as misplaced. 

\subsection{Performance of tag removal detection}
Similar to the evaluation of \textit{tag misplacement} detection, only the emulated dataset is utilized to evaluate the performance of \textit{tag removal} detection. As ground truth, tag 3, 10, 20, 30, 40 were actually removed in the meantime of data collection. Using the method described in section~\ref{sect:detectremoval}, the top-10 candidates of removed tags are $\{10, 40, 30, 21, 17, 3, 34, 23, 43, 13\}$ which successfully cover 4 out 5 removed tags. 3 removed tags will be successfully detected if top-5 candidates are flagged as removed.

\section{Summary and Future Work}
In this work, we abstracted three types of location-related attacks in indoor mobile crowdsensing. Additional location dependent fingerprints are collected as supplementary information for falsified data detection. Moreover, another two detection methods are proposed to detect \textit{tag misplacement} and \textit{tag removal}. In order to evaluate the proposed detection methods, both crowdsensed and emulated datasets are employed. Experiment results have shown that the proposed attack detection algorithms can indeed detect attacks with high accuracy. As a part of the future work, we would like to explore some more sophisticated detection algorithms for \textit{tag misplacement} and \textit{tag removal} and investigate other types of security threats in mobile crowdsourcing.

\balance
\bibliographystyle{abbrv}
\bibliography{mobibee,lightindoor}

\begin{thebibliography}{10}

\bibitem{aydin2014crowdsourcing}
B.~I. Aydin, Y.~S. Yilmaz, Y.~Li, Q.~Li, J.~Gao, and M.~Demirbas.
\newblock Crowdsourcing for multiple-choice question answering.
\newblock In {\em AAAI}, pages 2946--2953, 2014.

\bibitem{bouet2008rfid}
M.~Bouet and A.~L. Dos~Santos.
\newblock Rfid tags: Positioning principles and localization techniques.
\newblock In {\em Wireless Days, 2008. WD'08. 1st IFIP}, pages 1--5. IEEE,
  2008.

\bibitem{chen2015rise}
S.~Chen, M.~Li, K.~Ren, X.~Fu, and C.~Qiao.
\newblock Rise of the indoor crowd: Reconstruction of building interior view
  via mobile crowdsourcing.
\newblock In {\em Proceedings of the 13th ACM Conference on Embedded Networked
  Sensor Systems}, pages 59--71. ACM, 2015.

\bibitem{dong2009integrating}
X.~L. Dong, L.~Berti-Equille, and D.~Srivastava.
\newblock Integrating conflicting data: the role of source dependence.
\newblock {\em Proceedings of the VLDB Endowment}, 2(1):550--561, 2009.

\bibitem{dutta2009common}
P.~Dutta, P.~M. Aoki, N.~Kumar, A.~Mainwaring, C.~Myers, W.~Willett, and
  A.~Woodruff.
\newblock Common sense: participatory urban sensing using a network of handheld
  air quality monitors.
\newblock In {\em Proceedings of the 7th ACM conference on embedded networked
  sensor systems}, pages 349--350. ACM, 2009.

\bibitem{gao2014jigsaw}
R.~Gao, M.~Zhao, T.~Ye, F.~Ye, Y.~Wang, K.~Bian, T.~Wang, and X.~Li.
\newblock Jigsaw: Indoor floor plan reconstruction via mobile crowdsensing.
\newblock In {\em Proceedings of the 20th annual international conference on
  Mobile computing and networking}, pages 249--260. ACM, 2014.

\bibitem{khan2014and}
R.~Khan, S.~Zawoad, M.~M. Haque, and R.~Hasan.
\newblock ‘who, when, and where?’location proof assertion for mobile
  devices.
\newblock In {\em IFIP Annual Conference on Data and Applications Security and
  Privacy}, pages 146--162. Springer, 2014.

\bibitem{kim2013study}
E.~Kim, Y.-S. Lee, S.-Y. Kim, J.-W. Choi, and M.-S. Jung.
\newblock A study on the information protection modules for secure mobile
  payments.
\newblock In {\em IT Convergence and Security (ICITCS), 2013 International
  Conference on}, pages 1--2. IEEE, 2013.

\bibitem{langheinrich2009survey}
M.~Langheinrich.
\newblock A survey of rfid privacy approaches.
\newblock {\em Personal and Ubiquitous Computing}, 13(6):413--421, 2009.

\bibitem{lee2012qrloc}
M.~Lee and D.~Han.
\newblock Qrloc: User-involved calibration using quick response codes for wi-fi
  based indoor localization.
\newblock In {\em Computing and Convergence Technology (ICCCT), 2012 7th
  International Conference on}, pages 1460--1465. IEEE, 2012.

\bibitem{li2016survey}
Y.~Li, J.~Gao, C.~Meng, Q.~Li, L.~Su, B.~Zhao, W.~Fan, and J.~Han.
\newblock A survey on truth discovery.
\newblock {\em Acm Sigkdd Explorations Newsletter}, 17(2):1--16, 2016.

\bibitem{mohan2008nericell}
P.~Mohan, V.~N. Padmanabhan, and R.~Ramjee.
\newblock Nericell: rich monitoring of road and traffic conditions using mobile
  smartphones.
\newblock In {\em Proceedings of the 6th ACM conference on Embedded network
  sensor systems}, pages 323--336. ACM, 2008.

\bibitem{ni2004landmarc}
L.~M. Ni, Y.~Liu, Y.~C. Lau, and A.~P. Patil.
\newblock Landmarc: indoor location sensing using active rfid.
\newblock {\em Wireless networks}, 10(6):701--710, 2004.

\bibitem{ozdenizci2011development}
B.~Ozdenizci, K.~Ok, V.~Coskun, and M.~N. Aydin.
\newblock Development of an indoor navigation system using nfc technology.
\newblock In {\em Information and Computing (ICIC), 2011 Fourth International
  Conference on}, pages 11--14. IEEE, 2011.

\bibitem{pei2016survey}
L.~Pei, M.~Zhang, D.~Zou, R.~Chen, and Y.~Chen.
\newblock A survey of crowd sensing opportunistic signals for indoor
  localization.
\newblock {\em Mobile Information Systems}, 2016, 2016.

\bibitem{prince2012two}
G.~B. Prince and T.~D. Little.
\newblock {A two phase hybrid {RSS/AoA} algorithm for indoor device
  localization using visible light}.
\newblock In {\em Global Communications Conference (GLOBECOM)}, pages
  3347--3352. IEEE, 2012.

\bibitem{xuzheng16mobibee}
R.~Z. Qiang~Xu.
\newblock {MobiBee: A Mobile Treasure Hunt Game for Location-dependent
  Fingerprint Collection}.
\newblock In {\em Proceedings of the 2016 ACM conference on Pervasive and
  ubiquitous computing adjunct publication}. ACM, 2016.

\bibitem{xuzhengubicomp15}
{Qiang Xu, Rong Zheng, Steve Hranilovic}.
\newblock {IDyLL}: Indoor localization using inertial and light sensors on
  smartphones.
\newblock In {\em Proceedings of the 2015 ACM International Joint Conference on
  Pervasive and Ubiquitous Computing}, UbiComp '15, 2015.

\bibitem{Rai2012}
A.~Rai, K.~K. Chintalapudi, V.~N. Padmanabhan, and R.~Sen.
\newblock {Zee: zero-effort crowdsourcing for indoor localization}.
\newblock In {\em Proceedings of the 18th annual international conference on
  Mobile computing and networking}, pages 293--304. ACM, 2012.

\bibitem{sastry2003secure}
N.~Sastry, U.~Shankar, and D.~Wagner.
\newblock Secure verification of location claims.
\newblock In {\em Proceedings of the 2nd ACM workshop on Wireless security},
  pages 1--10. ACM, 2003.

\bibitem{Shen2013WIP}
G.~Shen, Z.~Chen, P.~Zhang, T.~Moscibroda, and Y.~Zhang.
\newblock Walkie-markie: Indoor pathway mapping made easy.
\newblock In {\em Proceedings of the 10th USENIX Conference on Networked
  Systems Design and Implementation}, nsdi'13, pages 85--98, Berkeley, CA, USA,
  2013. USENIX Association.

\bibitem{siira2009location}
E.~Siira, T.~Tuikka, and V.~Tormanen.
\newblock Location-based mobile wiki using nfc tag infrastructure.
\newblock In {\em Near Field Communication, 2009. NFC'09. First International
  Workshop on}, pages 56--60. IEEE, 2009.

\bibitem{xu2016mobibee}
Q.~Xu and R.~Zheng.
\newblock Mobibee: a mobile treasure hunt game for location-dependent
  fingerprint collection.
\newblock In {\em Proceedings of the 2016 ACM International Joint Conference on
  Pervasive and Ubiquitous Computing: Adjunct}, pages 1472--1477. ACM, 2016.

\bibitem{Yang2012}
Z.~Yang, C.~Wu, and Y.~Liu.
\newblock Locating in fingerprint space: wireless indoor localization with
  little human intervention.
\newblock In {\em Proceedings of the 18th annual international conference on
  Mobile computing and networking}, pages 269--280. ACM, 2012.

\bibitem{zhao2012bayesian}
B.~Zhao, B.~I. Rubinstein, J.~Gemmell, and J.~Han.
\newblock A bayesian approach to discovering truth from conflicting sources for
  data integration.
\newblock {\em Proceedings of the VLDB Endowment}, 5(6):550--561, 2012.

\end{thebibliography}
\end{document}